\documentclass[12pt,twocolumn]{article}

\usepackage[dvips]{graphics}
\begin{document}
\onecolumn
\pagestyle{empty}

\begin{center}
{\bf{\LARGE Mathematica: A System of Computer Programs}} \\
~\\
{\large Santanu K. Maiti$^*$} \\
~\\
{\em Physics and Applied Mathematics Unit, Indian Statistical
Institute, 203 Barrackpore Trunk Road, Kolkata-700 108, India}\\
~\\
{$^*$E-mail:\em santanu.maiti@isical.ac.in}

\end{center}

\noindent

\begin{center}
{\Large \bf Abstract}
\end{center}
Mathematica is a powerful application package for doing mathematics and
is used almost in all branches of science. It has widespread applications
ranging from quantum computation, statistical analysis, number theory,
zoology, astronomy, and many more. Mathematica gives a rich set of 
programming extensions to its end-user language, and it permits us to 
write programs in procedural, functional, or logic (rule-based) style, 
or a mixture of all three. For tasks requiring interfaces to the external 
environment, mathematica provides mathlink, which allows us to communicate 
mathematica programs with external programs written in C, C++, F77, F90, 
F95, Java, or other languages. It has also extensive capabilities for 
editing graphics, equations, text, etc. Starting from the basic level 
of mathematica here we illustrate how to use a mathematica notebook and 
write a program in the notebook. Following with this, we also describe 
very briefly about the importance of the local and global variables those 
are used in writing programs in mathematica. Next, we investigate elaborately 
the way of linking of external programs with mathematica, so-called the 
mathlink operation. Using this technique we can run very tedious jobs quite
efficiently, and the operations become extremely fast.
Sometimes it is quite desirable to run jobs in background of a computer
which can take considerable amount of time to finish, and this allows
us to do work on other tasks, while keeping the jobs running. The way 
of running jobs, written in a mathematica notebook, in background is 
quite different from the conventional methods i.e., the techniques for
the programs written in other languages like C, C$++$, F77, F90, F95, 
etc. To illustrate it, in the present article we study how to create a 
mathematica batch-file from a mathematica notebook and run it in 
the background. Finally, we explore the most significant issue of this
article. Here we describe the basic ideas for parallelizing a mathematica 
program by sharing its independent parts into all other remote computers 
available in the network. Doing the parallelization, we can perform large 
computational operations within a very short period of time, and therefore, 
the efficiency of the numerical works can be achieved. Parallel computation 
supports any version of mathematica and it also works significantly well 
even if different versions of mathematica are installed in different
computers. 
All the operations studied in this article run under any supported 
operating system like Unix, Windows, Macintosh, etc. For the sake of our
illustrations, here we concentrate all the discussions only for the Unix 
based operating system.

\newpage
\section{Introduction}

Mathematica, a system of computer programs, is a high-level computing
environment including computer algebra, graphics and programming. 
Mathematica is specially suitable for mathematics, since it incorporates
symbolic manipulation and automates many mathematical operations.
The key intellectual aspect of Mathematica is the invention of a new kind 
of symbolic computation language that can manipulate the very wide range of 
objects needed to achieve the generality required for technical computing
by using a very small number of basic primitives. Just a single line 
sometimes makes a meaningful program in mathematica--the syntax, documents 
and methodology used for input and output remaining as they are for immediate 
calculations. It supports every type of operation--be they data, functions, 
graphics, programs, or even complete documents--to be represented in a 
single, uniform way as a symbolic expression. This unification has many 
practical benefits to broadening the scope of applicability of each function. 
The raw algorithmic power of mathematica is magnified and its utility 
extended. Mathematica is now emerging as an important tool in many branches 
of computing, and today it stands as the world's best system for general 
computation.

Mathematica has widespread applications in different fields and is often 
used for research, loading and analyzing data, giving technical 
presentations and seminars etc. Mathematica is extraordinary well-rounded. 
It is suitable for both numeric and symbolic work, and it has remarkable 
word-processing capabilities as well. Mathematicians can search for a 
working model, do intensive calculation, and write a dissertation on the 
project (including complex graphics) -- all from within mathematica. It 
is mathematica's complete consistency in design at every stage that gives 
it this multilevel capability and helps advanced usage evolve naturally.

\section{Start with Mathematica}

We generally use mathematica through documents called {\em notebooks}.
To start a mathematica notebook in Unix we write `mathematica $\&$'
from a command line and then press the `Enter' key from the key-board.
A typical notebook consists of cells that may contain graphics, texts,
programs or calculations. Now to exit from a mathematica notebook
we first go to the command `File' and then press `Quit' from the menu
bar of the notebook. Without using a notebook one can also use mathematica
by typing the command `math' from a command line and all the jobs can
also be done as well. To exit from mathematica for this particular case,
we should write either `Exit' or `Quit' and then press the `Enter' key.
Thus one can run mathematica by using any one of the above two ways, but
the most general way to do the interactive calculations in mathematica
is the use of mathematica through notebook documents.

\subsection{Use of a Mathematica Notebook}

In a notebook, a job is performed in a particular cell and for different
jobs we use different cells. One can also use a single cell for all the
operations, but it is quite easy if different operations are performed in
separate cells. A cell is automatically created when we begin to write
anything in the notebook. After writing proper operation/operations, it
is needed to run the jobs. For this purpose, we press the key `Shift' and
holding this key, we then press `Enter' from the key-board. The results
for the inputs are evaluated and they are available immediately underneath
in a separate cell, so-called the output cell.

In mathematica we can do all kind of mathematical operations like numerical
computation, algebric computation, matrix manipulation, different types of
graphics etc., and all these things are clearly described by several examples
in key mathematica book of {\em Wolfram Research}~\cite{wolfram}. So in
this article we shall not give any such example further. Now to do large
numerical computations, it is needed to write a complete program. For this
purpose, here we describe something about the way of writing a complete
program in a mathematica notebook.

\subsection{Way of Writing a Program in Mathematica}

In mathematica, we can write a program efficiently compared to any other
existing languages. As illustrative example, here we mention a very simple
program which is: {\em the generation of a list of two random numbers and
the creation of a $2$D plot from these numbers}.

\vskip 0.1in
\noindent
The program is:

\vskip 0.2in
\begin{center}
{\fbox{\parbox{5.65in}{\centering{
sample$[$times$_-]$$:=$Block$[\{$local variables$\}$,

numbers$=$Table$[\{$Random$[]$, Random$[]$$\}$, $\{i$, $1$, times$\}]$;

figure$=$ListPlot$[$numbers, PlotJoined$\rightarrow$True,
AxesLabel$\rightarrow$$\{$xlabel, ylabel$\}$$]]$
}}}}
\end{center}

\vskip 0.12in
\noindent
This is the complete program for the generation of a list of two random
numbers and the creation of a $2$D plot from these numbers. This program
is written in a single cell. After the end of this program we run it
by using the command `SHIFT' $+$ `ENTER', and then the mathematica does
the proper operations and executes the result in an output cell.

Now to understand this program, it is necessary to describe the meaning
of the different commands used in this program. To start a program it is
necessary to specify a name for the particular program. In this case, we
specify `sample' as a program name, for the sake of simplicity. One can
also use any other name in place of `sample', since this is a dummy name.
If there is any running variable, like `times' (a dummy variable) in this
\begin{figure}[ht]
{\centering \resizebox*{10.0cm}{6.5cm}{\includegraphics{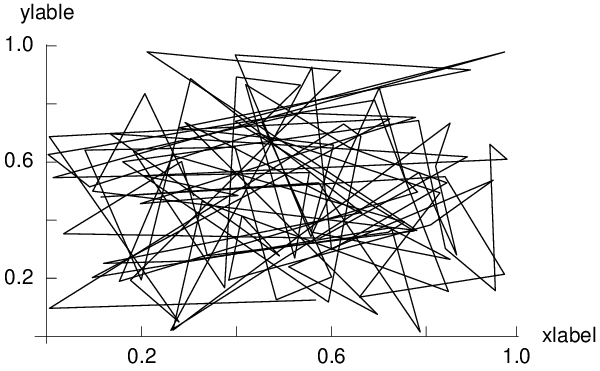}}\par}
\caption{A $2$D plot for a set of two random numbers.}
\label{random}
\end{figure}
particular case, then it has to be given within the bracket `$[~~]$'. After
that the symbol `$_-$' is used, which indicates the variable as a functional
variable. This is similar to define a functional variable, like $f[x]$ as
$f[x_-]$ in mathematica. Now all the mathematical commands those are used
for calculating the job are inserted within `Block$[\ldots ]$'. This is the
central part of the program. This portion i.e., `Block$[\ldots ]$' is
connected with `sample$[$times$_-]$' by the symbols `$:=$'. The symbol `$:$'
has an important role, and therefore it has to be taken into account
properly. Inside the `Block' the `local variables' for the program are
declared within the bracket `$\{~~\}$'. There may also exist another type
of variables called `global variables'. Later in this article, we will
focus about these two different types of variables in detail. Now the rest
part of the program differs from program to program depending on the
nature of the particular operations. In this program, first we construct a
list of two random numbers. In mathematica, a random number is generated
simply by using the command `Random$[]$'. Therefore, a list of two
such random numbers can be done very easily if we construct a table,
which is performed by the command `Table' as given in the program. The
integer $i$ runs from $1$ to `times', where the value of `times' can
be put anything. So if we write `sample$[10]$', here `times $=10$',
then $i$ runs from $1$ to $10$ and if we take `sample$[30]$', where
`times $=30$', then $i$ goes from $1$ to $30$. Now it becomes quite user
friendly if we mention different variable names for the different
mathematical operations which are {\em not exactly} identical with
any built in function available in mathematica like `Random', `Table',
`Plot', etc. In this program we use the variable names `numbers' and
`figure' for the two different operations. At the end of each mathematical
operation, except the last operation which gives the final output of a
program, we put the symbol `$;$'. This is also very crucial. Here we use
the symbol `$;$' at the end of the second line only, but not in the last
operation since this is the final output of this program. The command
`ListPlot' plots the list of data points where the command
`PlotJoined$\rightarrow$True' connects the lines between the data points.
Finally, the command `AxesLabel' in this line is an option for the graphics
functions to specify the labels in the axes.

The output for this program is shown in Fig.~\ref{random}, which appears
in a separate cell just below the input cell of the program. So now we
can easily write and compile a program in mathematica.

\subsubsection{Characterization of Local and Global Variables}

The local and global variables in mathematica play an important role, and
therefore care should be taken about these two types of variables when we
write a program in mathematica. We have already mentioned about the local
variables in the previous section that these variables are introduced only
inside the bracket `$\{~~\}$' at the beginning of the `Block$[~~]$'. In
such a case, the values of these parameters are only {\em defined within the
cell} where we write a particular program. Outside this cell, they are
undefined and therefore, we may also use these same parameters for writing
other programs without any trouble.

On the other hand, the global variables are those which are not used within
the bracket `$\{~~\}$' of a program. For such a case, these variables are
assigned throughout the notebook for all cells. Thus if we declare any value
for a such parameter, then it will read this particular value whenever we
use it in any program. Accordingly, it may cause a difficulty if we use the
same variable in another program by mistake. So we should take care about
these two types of variables. To make it clear, here we illustrate the
behavior of these two different kinds of variables by giving proper examples.

\vskip 0.2in
\begin{center}
{\fbox{\parbox{5.65in}{\centering{
sample$[$times$_-]$$:=$Block$[\{$$t=2.3,p=-1.5$$\}$,

numbers$=$Table$[\{$Random$[]$, Random$[]$$\}$, $\{i$, $1$, times$\}]$;

figure$=$ListPlot$[$numbers, PlotJoined$\rightarrow$True,
AxesLabel$\rightarrow$$\{$xlabel, ylabel$\}$$]]$
}}}}
\end{center}

\vskip 0.12in
\noindent
Let us consider the above program which is written in a particular cell in
the mathematica notebook. In this program, we introduce two local variables
$t=2.3$ and $p=-1.5$. Both these two variables are given inside the
bracket `$\{~~\}$'. Now if we check the values outside the cell then the
output will be simply $t$ and $p$ for these two variables. So these are
the local variables and one can safely use these parameters again in other
programs.

\vskip 0.2in
\begin{center}
{\fbox{\parbox{5.65in}{\centering{
sample$[$times$_-]$$:=$Block$[\{$$t=2.3,p=-1.5$$\}$,

$q=3.5$;

numbers$=$Table$[\{$Random$[]$, Random$[]$$\}$, $\{i$, $1$, times$\}]$;

figure$=$ListPlot$[$numbers, PlotJoined$\rightarrow$True,
AxesLabel$\rightarrow$$\{$xlabel, ylabel$\}$$]]$
}}}}
\end{center}

\vskip 0.12in
\noindent
Now we refer to this program where we introduce an extra line for
another variable $q=3.5$ compared to the previous program of this section.
Once we run this program, the value of $q$ will be assigned for any cell
of the notebook. Therefore, in this case $q$ becomes the global variable,
and if one uses it further in other program then the value of this parameter
$q$ will be assigned as $3.5$. Hence a mismatch will occur, and thus we
should be very careful about these two different types of parameters.

\section{Way to Link External Programs in Mathematica by Proper
         Math-Link Commands}

This section illustrates an important part of this article which
deals with the way of linking of an external program with mathematica
through proper mathlink commands. The mechanism for the linking of
external program written in C with mathematica has already been
established~\cite{maeder}. But this will not work if one tries to link
an external program written in other languages like F77, F90, F95, etc.,
with mathematica. This motivates us to find a way of linking an external
program written either in any one of these later languages (F77, F90, F95)
with mathematica. Here we illustrate it for the FORTRAN-90 source
files~\cite{smith,mart} only, but this mechanism will also work
significantly for the other Fortran source files as well.

\subsection{Mathlink for XL Fortran-90 Source Files}

In order to understand the basic mechanism for linking an external program
with mathematica, let us begin by giving a simple example. Here we set the
program as follows:

\begin{enumerate}
\item Construct two square matrices in mathematica.
\item Take the product of these two matrices by using an external program
written in F90.
\item Calculate the eigenvalues of the product matrix in mathematica.
\end{enumerate}

\begin{figure}[ht]
{\centering \resizebox*{10.0cm}{3.5cm}{\includegraphics{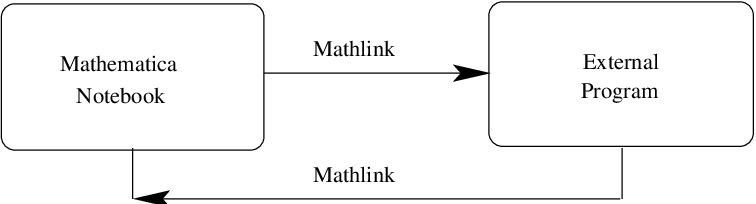}}\par}
\caption{Schematic representation of mathlink operations.}
\label{diagram}
\end{figure}

\vskip 0.12in
\noindent
The whole operations can be pictorially represented in Fig.~\ref{diagram}.

\vskip 0.12in
\noindent
The operations $1$ and $3$ are performed in mathematica, while the operation
$2$ is evaluated by the external program. The transformations of the datas
from the mathematica notebook to the external program are done by using
some proper commands, so-called mathlink operation. To complete this
particular job (operations $1$-$3$), we need two programs. One is written
in mathematica for the operations $1$ and $3$, while the other program is
written in F90 for the operation $2$. Now we describe all these steps one
by one. Let us first concentrate on the external program, given below,
where the multiplication of the two square matrices (operation $2$) is
performed. The first line of the program corresponds to the command line
where the symbol `!'
is used to make a statement as a command statement. The next line provides
a specific name of the program which is described by the command `program
multiplication'. This actually starts the program, and accordingly, the
program is ended by the command `end program multiplication'. In F90,
we can allocate and deallocate array variables in the programs which help
us a lot to save memory and are very essential to run many jobs
simultaneously. Here we use three array variables `a, b and c' for the
three different matrices whose dimensions are allocated by the order of
the matrix `n'. Finally, the product of the two matrices `a' and `b' is
determined by the command `matmul(a,b)' and the datas are stored in the
matrix `c'. This is the full program for the matrix multiplication of any
two square matrices of order `n'.

\vskip 0.2in
\begin{center}
{\fbox{\parbox{4.75in}{\centering{
! A program for matrix multiplication of two square matrices

program multiplication

implicit double precision (a-h,o-z)

double precision, allocatable :: a(:,:),b(:,:),c(:,:)

read * , n !(the order of the matrix)

allocate (a(n,n),b(n,n),c(n,n))

read * , ((a(i,j),j=1,n),i=1,n) ; read * , ((b(i,j),j=1,n),i=1,n)

!    Calculation of matrix multiplication :

c=matmul(a,b)

print'(1(1x,f10.6))',((c(i,j),j=1,n),i=1,n)

end program multiplication
}}}}
\end{center}

\subsubsection{Compilation and Optimization of XL Fortran-90 Source Files}

After writing a program, first we need to compile it to check whether
there is any syntax error or not to proceed for further operations.
Several commands are accessible for the compilation and optimization of
a program. The commands generally used to compile a F$90$ source file
are: xlf$90$, xlf$90_-$r, xlf$90_-$r$7$, etc. Thus we can use anyone of
these to compile this program, but different commands optimize a program
in different ways which solely depends on the nature of the particular
program. The simplest way for the compilation of a program is,

\vskip 0.2in
\begin{center}
{\fbox{\parbox{1.75in}{\centering{
xlf90 filename.f
}}}}
\end{center}

\vskip 0.12in
\noindent
With this operation, an `executable file' named `a.out' is created,
by default, in the present working directory (pwd). But if one uses several
programs simultaneously then it would be much better to specify different
names of different `executable files' for separate programs. To do this
we use the prescription,

\vskip 0.2in
\begin{center}
{\fbox{\parbox{2.55in}{\centering{
xlf90 filename.f -o filename
}}}}
\end{center}

\vskip 0.12in
\noindent
Under this process, the `executive file' named as `filename' is created.
Thus we can create proper `executive files' for different jobs and all
the jobs can be performed simultaneously without any difficulty.

For our illustrative purposes, below we mention some other optimization
techniques for the Fortran source files.

\begin{itemize}
\item -o : Optimizes code generated by the compiler.

\item -o0 : Performs no optimizations. (It is the same as -qnoopt.)

\item -o2 : Optimizes code (this is the same as -O).

\item -o3 : Performs the -O level optimizations and performs additional
      optimizations that are memory or compile time intensive.

\item -o4 : Aggressively optimizes the source program, trading off
      additional compile time for potential improvements in the
      generated code.  This option implies the following options:
     -qarch=auto -qtune=auto -qcache=auto -qhot -qipa.

\item -o5 : Same as -O4, but also implies the -qipa=level=2 option.
\end{itemize}

\vskip 0.12in
\noindent
From these operations, we can make some flavors about the compilation
and optimization technique for a Fortran source file. For a detailed
description of each operation, we refer to the XL Fortran User's
Guide~\cite{ibm}.

\subsubsection{Link of XL Fortran-90 Program with Mathematica}

This is the heart of this article. Below we set the mathematica program
for the operations $1$ and $3$, incorporating the operation $2$ by using
the proper mathlink commands, and illustrate all the steps properly
(Fig.~\ref{diagram}).

Let us suppose the external program, for the operation $2$, is
written in the directory `/allibmusers/santanu/files/test'. Generally we
are habituated to see the working directory as `/user/santanu/...' or
`/home/santanu/...' or `/allusers/santanu/...', etc. So it can be anything
like these. Thus knowing the directory where the external program is written,
we enter into that particular directory and compile the external program
properly to create an `executive file' for further operations. For this
particular case, we create the `executive file' named as `mat' which is
used in the $13$-th line of the following mathematica program. Now the
external program is ready for the operation, and we enter into the
directory where we will run the job in the mathematica notebook for the
operations $1$ and $3$.

Sitting in the directory where the mathematica notebook is open, we
need to connect the proper directory where the `executive file' for
the external program exists. The name of the pwd can be checked
directly from the mathematica notebook by using the command `Directory[]'.
Suppose the pwd is `/allibmusers/santanu/math'. Now If this pwd is
different from the directory where the file `mat' exists, then we make
a link to that particular directory through the command `SetDirectory'.
Below we give an example to connect the directory
`/allibmusers/santanu/files/test', where the file `mat' exists.

\vskip 0.2in
\begin{center}
{\fbox{\parbox{4in}{\centering{
SetDirectory[``/allibmusers/santanu/files/test"]
}}}}
\end{center}

\vskip 0.12in
\noindent
For this operation, the total path must be used within the double
quotes `` ". Using the command `ResetDirectory[]', we can come back to
the initial directory. Thus we can connect and disconnect any directory
with the pwd from the mathematica notebook, and able to link external
programs with mathematica very easily.

\vskip 0.2in
\begin{center}
{\fbox{\parbox{5.5in}{\centering{
sample$[$ns$_-]$$:=$Block$[\{$$t=0,s=0$,vacuum1$=\{\}$,vacuum2$=\{\}$$\}$,

SetDirectory[``/allibmusers/santanu/files/test"];

Do[Do[a1 = If[i == j, t, 1.213]; \\
    a2 = AppendTo[vacuum1, a1], $\{$j, 1, ns$\}$], $\{$i, 1, ns$\}$]; \\
mat1 = Partition[a2, ns];

Do[Do[a3 = If[k == l, s, 2.079]; \\
      a4 = AppendTo[vacuum2, a3], $\{$l, 1, ns$\}$], $\{$k, 1, ns$\}$]; \\
mat2 = Partition[a4, ns];

mat3 = Partition[Flatten[$\{\{$mat1$\}$, $\{$mat2$\}\}$], ns]; \\
matrixorder = $\{$ns$\}$; \\
output = Insert[mat3, matrixorder, 1]; \\
Export[``mat3.dat", output];

matrix = Partition[
   Flatten[ReadList[``$!$mat$<$mat3.dat", Number, \\
             RecordLists$\rightarrow$ True]], ns]; \\
  results = Eigenvalues[matrix]$]$
}}}}
\end{center}

\vskip 0.12in
\noindent
In the above program, the variables `t' and `s' are the local variables,
and we have already discussed about these variables in the previous section.
`vacuum1=$\{\}$' and `vacuum2=$\{\}$' are the two empty lists where the
datas are stored for each operation of the two `DO' loops given in the
program to make the lists `a2' and `a4' respectively. The `Partition'
command makes the partition of a list. The parameter `ns' gives the order
of the two square matrices. By using the command `Export' we send the file
`mat3.dat' which is treated as the input file for the external program
kept in the directory `/allibmusers/santanu/files/test'. To perform the
matrix multiplication by using the external program and get back the
product matrix in the mathematica notebook we use the operation:
ReadList[``!mat$<$mat3.dat", Number, RecordLists$\rightarrow$True].
Here the command `ReadLeast' is used to read the objects from a file
and the commands `Number' and `RecordLists$\rightarrow$True' are the
options of the command `ReadList'. Finally, the eigenvalues of the
matrix in the mathematica notebook are determined by using the
command `Eigenvalues'.

\subsubsection{Link of other XL Fortran Programs with Mathematica}

Now we can also use the mathlink operations for other programs written
either in F$77$ or F$95$ by the above mechanisms. For these programs, we
should use proper commands for the compilations and optimizations. As
representative example, here we mention some of the commands for the
compilation of these XL Fortran source files those are: xlf, f$77$,
fort$77$, xlf$_-$r, xlf$_-$r$7$, xlf$95$, xlf$95_-$r and xlf$95_-$r$7$.

So now we can able to use mathlink commands for any type of Fortran program.

\section{Way to Create a Mathematica Batch-file and Run it in Background}

In the above section (Sec. 3), we have studied in detail how to start
mathematica, write programs in mathematica and the way of linking of
external programs with a mathematica notebook by using proper mathlink
commands. Now it may be quite desirable to run jobs in background which 
take much time to finish, and to do other works in separate windows, 
keeping the jobs running. This motivates us to explore the basic 
mechanisms for running mathematica programs in background. It can be 
done by creating proper mathematica batch-file which we will describe 
here elaborately.

In order to understand the complete process, let us start by giving a very
simple example of a mathematica program. We set the program as follows: 

\vskip 0.15in
\noindent
{\em The generation of a list of two random numbers, a $2$D plot from
these set of random numbers and then the creation of an `EPS' file for
this 2D plot}.

\vskip 0.12in
\noindent
For this program, first we need to make a list of two random numbers
and then construct a 2D plot using this set of random numbers. Finally,
we make an `EPS' file for this plot. Here, we are mainly interested to 
run this complete job in background. Before doing this job in 
background, let us now describe the different mathematical operations 
with proper commands which are to be done in a mathematica notebook for 
this particular operation.

\vskip 0.1in
\noindent
The program for the generation of a set of two random numbers and a
2D plot from these numbers is as follows:

\vskip 0.2in
\begin{center}
{\fbox{\parbox{5.65in}{\centering{
sample$[$times$_-]$$:=$Block$[\{$local variables$\}$,

numbers$=$Table$[\{$Random$[]$, Random$[]$$\}$, $\{i$, $1$, times$\}]$;

fig$=$ListPlot$[$numbers, PlotJoined$\rightarrow$True,
AxesLabel$\rightarrow$$\{$xlabel, ylabel$\}$$]]$
}}}}
\end{center}

\vskip 0.12in
\noindent
To get the output of this program, we run it by entering some value for
the variable `times', like `sample[100]' or `sample[200]' etc. Then the
\begin{figure}[h]
{\centering \resizebox*{10.0cm}{6.5cm}{\includegraphics{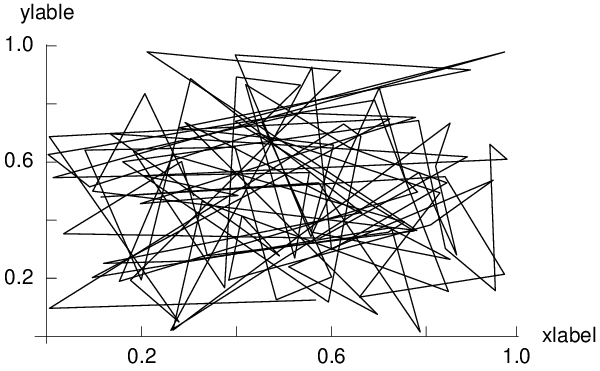}}\par}
\caption{A $2$D plot for a set of two random numbers.}
\label{random1}
\end{figure}
mathematica does the proper operations and executes the result in an
output cell. The output of the 2D plot is shown in Fig.~\ref{random1}.

\vskip 0.12in
\noindent
Now to create an `EPS' file for this 2D plot we use the following
operation:

\vskip 0.2in
\begin{center}
{\fbox{\parbox{3.15in}{\centering{
Export$[$``filename.eps", fig, ``EPS"$]$
}}}}
\end{center}

\vskip 0.12in
\noindent
In this above expression, the name `fig' is used to call the graphics
file, and the `eps' file is saved by the name `filename.eps' in the
present working directory.

Thus we are now clear about all the mathematical operations those
are to be done in a mathematica notebook for the above mentioned program.
So now we make our attention for running this program in background.

In order to run this program in background, first we need to create a
batch-file which is a text file from these mathematica input commands
those are written in different cells of a mathematica notebook. For this
purpose, we go through these steps:

(a) Select the cells from the mathematica notebook, and then follow the
direction by clicking on
{\em Cell} $\rightarrow$
{\em Cell Properties} $\rightarrow$ {\em Initialization Cell} from the menu
bar to initialize the cells.

(b) To generate the batch-file, follow the direction by clicking on
{\em File} $\rightarrow$ {\em Save As Special\ldots} $\rightarrow$
{\em Package Format} from the menu bar.

Then a dialog box appears for specifying the file name and the location
of the mathematica input file. Here we use the input file for the
operation of the mathematica job.

After these steps, let us suppose, we generate a batch-file named as
`santanu.m' for the above mathematica program. Generally the batch-files
for this purpose are specified by using the extension `.m' i.e., like
the name as `filename.m'. To run this batch-file `santanu.m' in background,
we use the following prescription:

\vskip 0.2in
\begin{center}
{\fbox{\parbox{4in}{\centering{
nohup time math $<$ santanu.m $>$ santanu.out $\&$
}}}}
\end{center}

The file name `santanu.out' is the output file, where all the outputs
for the different operations are available. To get both the input and
output lines of the mathematica notebook, it is necessary to use the
following command in the first line of the notebook.

\vskip 0.2in
\begin{center}
{\fbox{\parbox{2.35in}{\centering{
AppendTo[$\$$Echo, ``stdout"]
}}}}
\end{center}

At the end of all these steps, we get the output file `santanu.out' and
the graphics file `filename.eps' in `EPS' format in the present working
directory where the batch-file `santanu.m' is run in the background.

\section{Parallel Evaluation of Mathematica Programs}

Parallelization is a form of computation in which one can perform many
operations simultaneously. Parallel computation uses multiple processing
elements simultaneously to finish a particular job. This is accomplished
by breaking the job into independent parts so that each processing element
can execute its part of the algorithm simultaneously with the others. The
processing elements can be diverse and include resources such as a single
computer with multiple processors, several networked computers, specialized
hardware, or any combination of the above.

In this section, we narrate the basic mechanisms for parallelizing a
mathematica program by running its independent parts in several computers
available in the network. Since all the basic mathematical operations are
performed quite nicely in any version of mathematica, it does not matter
even if different versions of mathematica are installed in different computers
those are required for the parallel computing. 

\subsection{How to Open Mathematica Slaves in Local Computer ?}

In parallel computation, different segments of a job are computed 
simultaneously. These operations can be performed either in a local 
computer or in remote computers available in the network. Separate 
operations are exhibited in separate mathematica slaves. In order to
emphasize the basic mechanisms, let us now describe the way of starting
a mathematica slave in a local computer. To do this, first we load the
following package in a mathematica notebook.

\vskip 0.2in
\begin{center}
{\fbox{\parbox{2.25in}{\centering{
Needs[``Parallel\`{}Parallel\`{}"]
}}}}
\end{center}
To enable optional features, then we load the package,

\vskip 0.2in
\begin{center}
{\fbox{\parbox{2.45in}{\centering{
Needs[``Parallel\`{}Commands\`{}"]
}}}}
\end{center}
Now we can open a mathematica slave in the local computer by using the 
command,

\vskip 0.2in
\begin{center}
{\fbox{\parbox{4.25in}{\centering{
LaunchSlave[``localhost", ``math -noinit -mathlink"]
}}}}
\end{center}
Using this command, several mathematica slaves can be started from the
master slave. Now it becomes much more significant if we specify the 
names of different slaves so that independent parts of a job can be shared
into different slaves appropriately. For our illustrations, below we
give some examples how different slaves can be started with specific 
names.

\vskip 0.2in
\begin{center}
{\fbox{\parbox{4.25in}{\centering{
link1=LaunchSlave[``localhost", ``math -noinit -mathlink"] \\
link2=LaunchSlave[``localhost", ``math -noinit -mathlink"] \\
link3=LaunchSlave[``localhost", ``math -noinit -mathlink"] 
}}}}
\end{center}
Here link1, link2 and link3 correspond to the three different slaves. 
The details of these slaves can be available by using the following 
command,

\vskip 0.2in
\begin{center}
{\fbox{\parbox{5.25in}{\centering{
TableForm[RemoteEvaluate[$\{\$$ProcessorID, $\$$MachineName, $\$$SystemID,
$\$$ProcessID, $\$$Version$\}$], TableHeadings$\rightarrow$$\{$None,$\{$``ID",
``host", ``OS", ``process", ``Mathematica Version"$\}\}$]
}}}}
\end{center}
The output of the above command becomes (as an example),
\vskip 0.2in
\begin{center}
\begin{tabular}{|c|c|c|c|c|}
\hline
ID & host & OS & process & Version \\ \hline 
1 & tcmpibm & AIX-Power64 & 463002 & 5.0 for IBM AIX Power (64 bit) \\ 
 & & & & (November 26, 2003) \\
2 & tcmpibm & AIX-Power64 & 299056 & 5.0 for IBM AIX Power (64 bit) \\ 
 & & & & (November 26, 2003) \\
3 & tcmpibm & AIX-Power64 & 385182 & 5.0 for IBM AIX Power (64 bit) \\ 
 & & & & (November 26, 2003) \\
\hline
\end{tabular}
\end{center}
The results shown in this table are for the above three slaves named as
link1, link2 and link3 respectively, where all these slaves are opened
from the local computer named as `tcmpibm' (say). To get the information
about the total number of slaves those are opened, we use the command,

\vskip 0.2in
\begin{center}
{\fbox{\parbox{1.45in}{\centering{
Length[$\$$Slaves]
}}}}
\end{center}
For this case, the total number of slaves becomes $3$.

\subsection{How to Open Mathematica Slaves in Remote Computers Available 
in Network ?}

To start a slave in remote computer, the command `ssh' is used which 
offers secure cryptographic authentication and encryption of the 
communication between the local and remote computer. Before starting 
a slave in a remote computer, it is necessary to check whether `ssh'
is properly configured or not, and this can be done by using the 
prescription,

\vskip 0.2in
\begin{center}
{\fbox{\parbox{1.75in}{\centering{
ssh remotehost math
}}}}
\end{center}
For example, if we want to connect a remote computer named as `tcmpxeon',
we should follow the command as,

\vskip 0.2in
\begin{center}
{\fbox{\parbox{1.75in}{\centering{
ssh tcmpxeon math
}}}}
\end{center}
Since `ssh' connection for a remote computer is password protected, it 
is needed to insert proper password, and if `ssh' is configured correctly,
the above operation shows the command `In[1]:='. Once `ssh' works 
correctly, a mathematica slave can be opened in a remote computer 
through this command,

\vskip 0.2in
\begin{center}
{\fbox{\parbox{4.75in}{\centering{
LaunchSlave[``remotehost", ``ssh -e none \`{}1\`{} math -mathlink"]
}}}}
\end{center}
For our illustrative purposes, below we describe how different slaves
with proper names can be started in different remote computers.

\vskip 0.2in
\begin{center}
{\fbox{\parbox{5.65in}{\centering{
link1=LaunchSlave[``tcmpxeon.saha.ac.in", ``ssh -e none \`{}1\`{} 
math -mathlink"] \\
link2=LaunchSlave[``tcmp441d.saha.ac.in", ``ssh -e none \`{}1\`{} 
math -mathlink"] \\
link3=LaunchSlave[``tcmpxeon.saha.ac.in", ``ssh -e none \`{}1\`{} 
math -mathlink"] \\
link4=LaunchSlave[``tcmp441d.saha.ac.in", ``ssh -e none \`{}1\`{} 
math -mathlink"] \\
}}}}
\end{center}
Here link1, link2, link3 and link4 are the four different slaves, where
the link1 and link3 are opened in a remote computer named as `tcmpxeon'
(say), while the other two slaves are started in another one remote
computer named as `tcmp441d' (say). Using this prescription, several
mathematica slaves can be started in different remote computers available 
in the network. The details of the above four slaves can be expressed
in the tabular form as,

\vskip 0.2in
\begin{center}
\begin{tabular}{|c|c|c|c|c|}
\hline
ID & host & OS & process & Version \\ \hline 
1 & tcmpxeon & Linux & 5137 & 5.0 for Linux (November 18, 2003) \\ 
2 & tcmp441d & Linux & 11323 & 5.0 for Linux (November 18, 2003) \\ 
3 & tcmpxeon & Linux & 5221 & 5.0 for Linux (November 18, 2003) \\ 
4 & tcmp441d & Linux & 11368 & 5.0 for Linux (November 18, 2003) \\ 
\hline
\end{tabular}
\end{center}
Thus we are now able to start mathematica slaves in local computer
as well as in remote computers available in the network, and with
this above background, we can describe the mechanisms for parallelizing
a mathematica program.

\subsection{Parallelizing of Mathematica Programs by using Remote Computers
Available in Network}

In order to understand the basic mechanisms of parallelizing a mathematica
program, let us begin with a very simple problem. We set the problem as
follows:

\vskip 0.15in
\noindent
{\bf\em\underline{Problem}}: {\em Construct a square matrix of any order
in a local computer and two other square matrices of the same order with
the previous one in two different remote computers. From the local computer,
read these two matrices those are constructed in the two remote computers.
Finally, take the product of these three matrices and calculate the
eigenvalues of the product matrix in the local computer.}

To solve this problem we proceed through these steps in a mathematica 
notebook. 

\vskip 0.15in
\noindent
{\em\underline{Step-1}} : For the sake of simplicity, let us first define
the names of the three different computers those are needed to solve this
problem. The local computer is named as `tcmpibm', while the names of the
other two remote computers are as `tcmpxeon' and `tcmp441d' respectively.
Opening a mathematica notebook in the local computer, let us first load
the package for parallelization, and to get the optional features, we load
another one package as mentioned earlier in Section 2. Then we start two 
mathematica slaves named as `link1' and `link2' in the two remote computers
`tcmpxeon' and `tcmp441d' respectively by using the proper commands as 
discussed in Section 3.

\vskip 0.15in
\noindent
{\em\underline{Step-2}} : Next we make ready three programs for the three
separate square matrices of same order in the local computer. Out of which 
one program will run in the local computer, while the rest two will run in
the two remote computers. These three programs are as follows.

\vskip 0.2in
\begin{center}
I. {\fbox{\parbox{5in}{\centering{
sample1$[$ns$_-]$$:=$Block$[\{$esi$=0, t=1.2, p=2.1$,vacuum1$=$$\{\}$$\}$,

Do[Do[a1=If[i$==$j,esi,0];

a2=If[$i<j$ $\&\&$ Abs$[i - j]$$==1, t, 0$];

a3=If[$i>j$ $\&\&$ Abs$[i - j]$$==1, p, 0$];

a4=a1+a2+a3;

a5 = AppendTo[vacuum1,a4], $\{j, 1, ns\}$], $\{i, 1, ns\}$];

a6 = Partition[a5, ns]$]$
}}}}
\end{center}

\vskip 0.2in
\begin{center}
II. {\fbox{\parbox{5in}{\centering{
sample2$[$ns$_-]$$:=$Block$[\{$esi$=0, q=2.6, r=1.8$,vacuum2$=$$\{\}$$\}$,

Do[Do[a1=If[i$==$j,esi,0];

a2=If[$i<j$ $\&\&$ Abs$[i - j]$$==1, q, 0$];

a3=If[$i>j$ $\&\&$ Abs$[i - j]$$==1, r, 0$];

a4=a1+a2+a3;

a5 = AppendTo[vacuum2,a4], $\{j, 1, ns\}$], $\{i, 1, ns\}$];

a6 = Partition[a5, ns]$]$
}}}}
\end{center}

\vskip 0.2in
\begin{center}
III. {\fbox{\parbox{5in}{\centering{
sample3$[$ns$_-]$$:=$Block$[\{$esi$=0, u=2, v=3$,vacuum3$=$$\{\}$$\}$,

Do[Do[a1=If[i$==$j,esi,0];

a2=If[$i<j$ $\&\&$ Abs$[i - j]$$==1, u, 0$];

a3=If[$i>j$ $\&\&$ Abs$[i - j]$$==1, v, 0$];

a4=a1+a2+a3;

a5 = AppendTo[vacuum3,a4], $\{j, 1, ns\}$], $\{i, 1, ns\}$];

a6 = Partition[a5, ns]$]$
}}}}
\end{center}
Since we are quite familiar about the way of writing mathematica 
programs~\cite{wolfram}, we do not describe here the meaning of 
the different symbols used in the above three programs further. Thus 
by using these programs, we can construct three separate square 
matrices of order `ns'.

\vskip 0.15in
\noindent
{\em\underline{Step-3}} : We are quite at the end of our complete operation.
For the sake of simplicity, we assume that, the program-I is evaluated in
the local computer, while the program-II and program-III are evaluated in
the two remote computers respectively. All these three programs run 
simultaneously in three different computers. To understand the basic
mechanisms, let us follow the program.

\vskip 0.2in
\begin{center}
{\fbox{\parbox{5.75in}{\centering{
sample4$[$ns$_-]$$:=$Block$[\{\}$,

ExportEnvironment[``Global\`{}"];

mat1 = sample1[ns];

RemoteEvaluate[Export[``data1.dat", sample2[ns]], link1];

RemoteEvaluate[Export[``data2.dat", sample3[ns]], link2];

mat2 = RemoteEvaluate[ReadList[``data1.dat", Number, RecordLists$\rightarrow$ 
      True], link1];

mat3 = RemoteEvaluate[ReadList[``data2.dat", Number, RecordLists$\rightarrow$ 
      True], link2];

mat4 = mat1.mat2.mat3;

Chop[Eigenvalues[mat4]]$]$

}}}}
\end{center}
This is the final program. When it runs in the local computer, one matrix
\begin{figure}[ht]
{\centering \resizebox*{10.0cm}{6.5cm}{\includegraphics{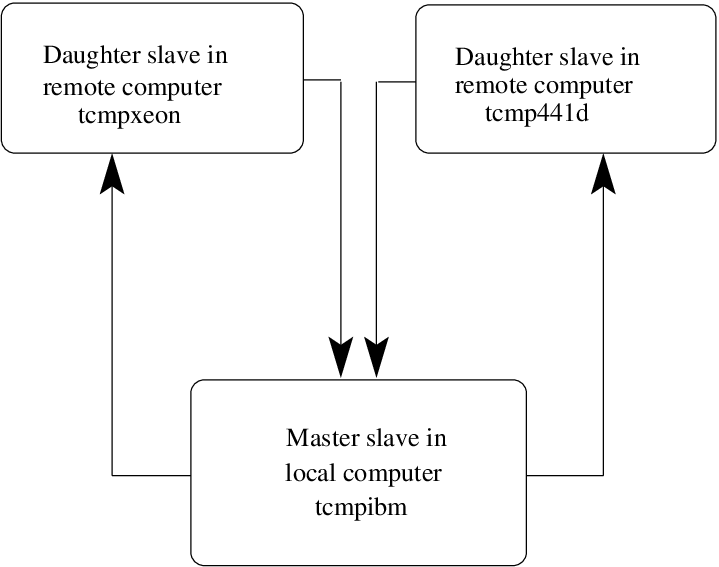}}\par}
\caption{Schematic representation of parallelization.}
\label{pic1}
\end{figure}
called as `mat1' is evaluated in the local computer ($3$rd line of the 
program), and the other two matrices are determined in the remote computers
by using the operations given in the $4$th and $5$th lines of the program
respectively. The $2$nd line of the program gives the command for the 
transformations of all the symbols and definitions to the remote computers.
After the completion of the operations in remote computers, we call back
these two matrices in the local computer by using the command `ReadList', 
and store them in `mat2' and `mat3' respectively. Finally, we take the
product of these three matrices and calculate the eigenvalues of the 
product matrix in the local computer by using the rest operations of the
above program.

\vskip 0.12in
\noindent
The whole operations can be pictorially represented in Fig.~\ref{pic1}.

\vskip 0.12in
\noindent
At the end of all the operations, we close all the mathematica slaves by
using the following command.

\vskip 0.2in
\begin{center}
{\fbox{\parbox{1.25in}{\centering{
CloseSlaves[  ]
}}}}
\end{center}

\noindent
\addcontentsline{toc}{section}{\bf {Concluding Remarks}}
\begin{flushleft}
{\Large \bf {Concluding Remarks}}
\end{flushleft}
\vskip 0.1in
\noindent
In summary, the basic operations presented in this communication may
be quite helpful for the beginners. Starting from the basic level,
in Section $2$, we have explored how to start mathematica, open a
mathematica notebook, write a program in mathematica, etc. Following 
with this, we have also described the utilities of the local and global 
variables those are used for writing programs in mathematica.

Later, in Section $3$, we have illustrated the basic mechanisms for the 
linking of external programs with mathematica notebook. This mathlink 
operation is an important part of this article, and it is extremely 
crucial for doing large numerical computations. Here we have concentrated 
the mathlink operation mainly for the XL Fortran 90 source files. But 
this operation can also be used for any other Fortran source file. 
In this section, we have also illustrated very briefly about the 
optimization techniques for the Fortran source files which may help us 
to run very complicated jobs quite efficiently.

In Section $4$, we have addressed in detail how to set up a mathematica
batch-file from a mathematica notebook and run it in the background
of a computer. Several programs are there which can take a considerable
amount of time to run. Some may take few days or even few weeks to
complete their analysis. For this reason, it may be desirable to place
such jobs in the background. This is a way of running a program that
allows one to continue working on other tasks (or even log out) while
still keeping the program running. Furthermore, backgrounded jobs are
not dependent on our session remaining open, so even if our computer
crashes, the job will continue uninterrupted.

At the end, in Section $5$, we have explored the basic mechanisms for 
parallelizing a mathematica program by running its independent parts in 
remote computers available in the network. By using this parallelization 
technique, one can enhance the efficiency of the numerical works, and it 
helps us to perform all the mathematical operations within a very short 
period of time. 

Throughout the article, we have focused all the basic operations for 
the Unix based operating system only. But all these operations also work 
very well in any other supported operating system like Windows, Macintosh, 
etc. 

\noindent
\addcontentsline{toc}{section}{\bf {Acknowledgment}}
\begin{flushleft}
{\Large \bf {Acknowledgment}}
\end{flushleft}
\vskip 0.1in
\noindent
I acknowledge with deep sense of gratitude the illuminating comments and
suggestions I have received from Prof. Sachindra Nath Karmakar during the
preparation of this article.

\addcontentsline{toc}{section}{\bf {References}}


\begin{thebibliography}{99}

\bibitem{wolfram} Stephen Wolfram. {\em Mathematica-5.0}.
\bibitem{maeder} Roman E. Maeder. {\em About Parallel Computing Toolkit}.
         A Wolfram Research Application Package.
\bibitem{smith} I. M. Smith. {\em Programming in Fortran 90 : A First Course
         for Engineers and Scientists}. University of Manchester, UK.
\bibitem{mart} Martin Counihan. {\em Fortran 90}. University of Southampton.
\bibitem{ibm} IBM. {\em XL Fortran for AIX : User's Guide}.

\end{thebibliography}
\end{document}